# Security in Carrier Class Server Applications for All-IP Networks

Marc Chatel* (Marc.Chatel@ericsson.com),
Michel Dagenais** (Michel.Dagenais@polymtl.ca),
Charles Levert* (Charles.Levert@ericsson.com),
and Makan Pourzandi*[1] (Makan.Pourzandi@ericsson.com)

*Open Systems Laboratory, Ericsson Research Canada
8400 Décarie Blvd.
Town of Mount Royal, Quebec
Canada, H4P 2N2

**Dept. of Computer Engineering
École Polytechnique de Montréal
C.P. 6079, Succ. CV, Montreal, Quebec
Canada, H3C 3A7

## Abstract

A revolution is taking place in telecommunication networks. New services are appearing on platforms such as third generation cellular phones (3G) and broadband Internet access. This motivates the transition from mostly switched to all-IP networks. The replacement of the traditional shallow and well-defined interface to telephony networks brings accrued flexibility, but also makes the network accordingly difficult to properly secure. This paper surveys the implications of this transition on security issues in telecom applications. It does not give an exhaustive list of security tools or security protocols. Its goal is rather to initiate the reader to the security issues brought to carrier class servers by this revolution.

## Introduction

Telephony networks typically interface to subscribers through a shallow and well-defined interface. After the introduction of out-of-band control technology based on Signaling

---

[1] Contact person.

February 2002

System 7 (SS7) in 1970s, the number of security incidents involving the core telephone network infrastructure reduced substantially. From then on, the commands that subscribers could send were limited to tone or pulse dialing of digits for signaling, switch hook flashes for simple features like call waiting and 3-way calling, and some dial access codes for features like caller-id blocking. On the subsequent incidents, never was the telephony core infrastructure compromised.

The telephony operating environment, however, is experiencing dramatic changes. Companies previously labeled as telephone operators are now offering broadband data access and numerous IP services, including mail and web hosting. Furthermore, these services may not remain totally separate from the traditional telephony channels, as more closely new integrated services are offered:

- Third generation cellular phones offer voice and high-speed data communications.

- Location services enable applications to query the precise location of cellular phones for emergency response, or targeted information/advertisement purposes.

- Many of the data-oriented applications being deployed are directly derived from popular Internet applications, or give direct access to Internet-located information content.

The result is the current move toward all-IP and IP-interoperable networks (Figure 1). The resulting communication infrastructure, integrating voice, data and multimedia, can be considered as a part of the single large global network, the Internet. It contains traditional wired and wireless phones and computers, and increasingly multi-functional small computers presented as telephony enabled personal digital assistants (PDA). The stateless phone of yesterday is replaced by a small computer, which is both vulnerable to attacks and capable of launching attacks. With the Internet, the explosion of the communication network brought a new field of possible threats [SANS2001, CERT2001]: attacks on or through the communication infrastructure between the server and the clients. The challenge is to have a network offering the flexibility associated with the Internet, while preserving the security and reliability expected from carrier grade equipment.

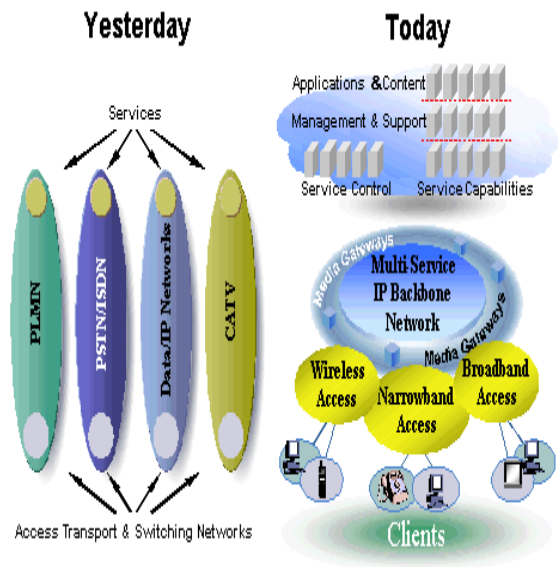

**Figure 1:** In the next generation telecom systems, the situation changes from vertical markets to an all-IP network with multiple service providers.

In security, there is the notion of an enemy that perpetrates systematic attacks against a system. Defending against systematic attacks contrasts with defending against undesirable events that occur at random. This is a fundamental change from the traditional telecom approach concentrated around countering undesirable events, rather than systematic, deliberate and repetitive attacks from different sources. Defending against random events can be accomplished by using techniques, e.g., redundancy, that tilt the odds in one's favor. Redundancy is useless against a systematic attack since it just adds the requirement that the very same attack be performed twice or more times. Defending against systematic attacks requires using specific countermeasures that make things impossible or at least very impractical for attackers. Words such as robust and reliable have different meanings in both contexts and must be interpreted accordingly; one kind of robustness does not imply the other.

In Section 2, the security threats are reviewed, while Section 3 examines protection measures associated with IP networks. Section 4 discusses the specific requirements of carrier class environments, and the relative importance in this context of the different security threats from Section 2 and security measures discussed in Section 3.

# Security Threats

An attack is an attempt to circumvent the security policy of a system, to make it do something it is not intended to do. When a possible attack is considered in context, it becomes a threat and can be assigned a risk factor (i.e., a probability of occurrence or an expectation of damage) that essentially serves to prioritize this type of attack. This context includes the known or estimated vulnerabilities that are present on a system and the measures that are put in place to counter their exploitation by specific attacks. Threats against a system are enumerated in a threat model that is the basis for developing a customized security policy, or strategy. One actual instance of an attack is referred to as an incident.

Since a system can be compromised with just one breach, easiest attacks (the weakest links of the chain) must be the first ones to be countered.

## Threat Model

For every threat, can be defined:

- Vulnerability exploited

- Countermeasures

- Likelihood

- Harm: it explains the type of the damage to the system, for example the loss of hard disk content.

- Impact: it describes the seriousness of the attack, major or minor damages,

- Risk: it is a function of the likelihood of a threat, the resulting impact, and the effectiveness of countermeasures to it. The table below presents the risk level according to the likelihood of a threat and its impact.

|  | Low likelihood | Medium likelihood | High likelihood |
|---|---|---|---|
| Low impact | Low risk | Medium risk | High risk |
| Medium impact | Low risk | Medium risk | High risk |
| High impact | Medium risk | High risk | Critical risk |

# Types of Attacks

It is possible to classify attacks according to different factors. For example:

- Based on the techniques that are used and the vulnerabilities that are exploited [Ranum1996].

- By the kind of gain they give the attacker upon success relative to the position held by the attacker prior to the attack.

- Remote and local attacks can be stringed together in a process known as privilege escalation.

- Preliminary attacks [Scambray2001] can assist a main attack, e.g., by making sure the attacker remains unidentified.

- Based on the fact that origin of the attack is in internal to the corporate network or external to it.

In the context of a telecom infrastructure, the possibility of local attacks must especially not be neglected. Indeed, their local systems are not limited to a single switch and may be quite large and complex; they are typically clusters of resources that are connected by internal protocols. They also provide access to different users and processes with different needs that should translate into different privileges.

If we classify possible types of attacks in terms of what are the attackers' motivations, we can list the following attacks:

- Denial of service: bringing computing or networking resources to exhaustion so that a service is unavailable, as long as the attack lasts, or the host or network element crashes (until manual intervention or automatic re-initialization).

- Theft of service: obtaining a service for which one does not subscribe, or a better grade of service (e.g., gold instead of silver grade QoS in DiffServ) than the one paid for.

- Information theft: Unauthorized copying of information. This is typically done for information that can be used for financial gain (such as credit-card numbers), but can be done for other purposes, such as software piracy.

- "Socially motivated" attacks: an attacker may wish to impersonate somebody specific in order to commit detectable crimes and have the other person charged. Attackers may also wish to target individuals or organizations as a matter of personal revenge, or as part of a violent activist or terrorist activity.

These attacks can be performed using different mechanisms:

- Social engineering: fooling someone, with minimal or no use of technology.

- Masquerading (or impersonation): breaching authentication (which can be based on secrets, tokens, or biometrics), for instance by theft or eavesdropping.

- Exploit of implementation flaw: purposely stimulating a system that fails to properly validate or manipulate external inputs of all kinds in a wrong way. This covers memory overflows, format string vulnerabilities, and improper decoding of URL % syntax and UTF-8.

- Data driven attacks: using Trojans and viruses.

- Network infrastructure attacks: exploiting design flaws in the protocols that implement the Internet infrastructure, for example: DNS spoofing, source routing, routing tables manipulations, generation of many ICMP replies by sending an ICMP request to a broadcast address, TCP connections termination or hijacking, and using FTP to order one server to inject data on a specific port of another server.

# Discussion

Every attack must be countered, or, at least, countermeasures must be described so that their use may be evaluated. Some of the attacks that have been exposed are made possible by fundamental elements of Internet design that one may have to live with. Some attacks can be countered by simply making an effort to implement and/or configure existing security-relevant solutions.

The source of many vulnerabilities on the server side is faulty software or misconfigurations. Attacks linked to software reliability can be countered by attempting to improve the reliability and by responding rapidly when vulnerabilities are exposed. Attacks linked to misconfigurations still can be countered by proper training and the definition of clear procedures.

# The Field of Security

Security is part of a bigger picture called risk management [Schneier2000]. Some of the risks involved are not linked to the business model or the market situation, but rather to deliberate undesirable events meant to harm the company. Although the implicit reaction would be to systematically fight back against such events, doing so costs money and there are actually several sensible options to consider. One possibility is simply to take the hit and accept the costs associated with an incident. Another is to purchase insurance against the associated threat. The final option is to fight back; this is what the field of security is all about. However, it is important to keep in mind that, in order to first choose an option, an analysis of costs/benefits (expected losses caused by the attack, cost of countermeasures), and risks (probability of successful attack with and without the countermeasures), has to be performed. In any case, a contingency and recovery plan is needed.

There are several types of security interventions: preventive measures, detection measures, response measures, recovery measures and forensics. No amount of prevention can make all undesirable events impossible; doing so would cost an infinite amount of money. A given level of prevention can thus be seen as a way to buy time before detection and response measures are performed, and as damage control to minimize the damages from a possible attack. Detection is an important yet very often neglected part of security activity. No response can be performed without prior detection, though the detection measures should collect the right information to prepare for an adequate response.. Despite all the preventive measures, there is no 100% secure system, therefore the recovery measures should be thought of and be in place at the time of setting up any system. Finally, forensics measures are primordial to find the perpetrators and prepare the company in face of possible legal issues.

## Main Properties

In security, three main properties are usually considered [Garfinkel1999, Russell1991]:

- Confidentiality (or secrecy): A secure computer must not allow information to be disclosed to any user or process not authorized to access it.

- Integrity (or accuracy): The system must not corrupt the information or allow any unauthorized malicious or accidental change. In network communications, a variant of accuracy is called authenticity. It provides a way to verify the origin of data by determining who entered or sent the data.

- Availability: The computer system's hardware and software must keep working efficiently and the system must be able to recover quickly and completely if a disaster occurs.

    The above definitions are often completed by [Anderson2001, Ryan2003]:

- Accountability: A secure computer should provide with the evidence on the actions of the users so that they can not successfully deny at a later time those actions

- Privacy: The computer should protect the personal secrets and information related to different users.

These properties have traditionally been emphasized in that order (CIA) in the intelligence community. In the contemporary context of the Internet, emphasis in the reverse order is more pertinent (AIC); these three are the ones most widely accepted in the literature. Though, other related properties have also been mentioned.

## Standards

From a telecommunications industry standpoint, national boundaries now have less influence on infrastructure deployment than in the past. Because of this, the necessity of an internationally accepted standard for security becomes more and more important.

Past efforts of the American Trusted Computer System Evaluation Criteria (TCSEC), or "Orange Book", and the European Information Technology Security Evaluation Criteria (ITSEC) have now been merged as the Common Criteria (CC), also known as ISO standard 15408. Such a standard introduces concepts that include evaluation, protection profiles, assurance components, and assurance levels [CSE1999].

## Context

The context in which the security mechanisms should be implemented is very important. Context parameters include:

- Hardware context: hardware hosting server and client applications. The factors to take into account are: processing power, available memory

- Transport Media: connection to outside world, network infrastructure. For example: LAN/WAN, all-IP-network, radio waves (more susceptible to eavesdropping), etc.

- Software:

    o OS: irrespective of the software mechanisms built at application level, the node's OS plays an essential role in efficiency and feasability of security mechanisms.

    o Middleware: it plays an important role in the development of the global information infrastructure. Middleware based service architectures facilitate the rapid development and deployment of new services by enabling software reuse and hiding the complexity of programming distributed applications.

- Application: Robustness for server and client applications plays an essential role in the security of the system.

## Hardware Security

Flaws in hardware can be used to eavesdrop on essential data like private keys used for cryptography. These flaws can be a plant error or a deliberate back door. An adequate audit of hardware material by the buyer, or buying reports from third party companies, are solutions.

As encryption/decryption of outgoing and ingoing data becomes inevitable tasks in new systems, there will be more and more specialized hardware available. Many big companies have ongoing plans for integrating specialized circuits to their commercial processors. The availability of these specialized hardware and their ever-decreasing costs, combined with the increasing costs of losses due to security incidents, will ease their penetration and use in the market.

## Software Security

Software is a fundamental part of the server. Unfortunately, most security vulnerabilities are software related.

### Operating System

The following operating systems are suitable to be used for telecommunication class applications. They include Unix-like operating systems such as Linux, FreeBSD, OpenBSD, Solaris, and proprietary operating systems developed by Telecom companies.

An operating system, by its architecture, can simplify or complicate security policy implementation. For example, it is obvious that an operating system that does not support any access control is much more open to security holes on the application or middleware levels than an operating system providing developers with strict access controls.

One important aspect of software security is that for security mechanisms to be efficient, they must be unbypassable. The operating system is therefore the best place to implement these mechanisms.

### Middleware

The middleware software must provide the developers with means to implement security mechanisms transparently and efficiently. As the lifetime for a middleware software is generally longer than an application, it makes more economical sense to concentrate efforts on making this layer secure. Also, it is more likely that security holes will be identified over time and solved.

However, having a middleware software providing security services is necessary, but it is not enough to make an application secure. At the application level, security middleware must provide developers with:

- Access controls: for application's functions and data, at sufficient granularity

- Audit of application events

- Authorization: Control of secure invocations according to the user's privileges

- Authentication: Use credentials to identify initiating principal

- Non-repudiation: Generate/verify evidence of actions, e.g., proof of creation receipt, origin, delivery of data. Evidence includes information to prove integrity of data, date/time, origin, action,

- Logging: Application transfers/stores evidence. It must be possible for developers to decide what to log.

These mechanisms can be explicitly implemented through calls to an API, or calls to services (or agents) of middleware. They can also be implicitly implemented through activation of security services. For example, in object-oriented design, one can imagine having secure objects, which log all their method invocations, used as ancestor classes transparently to the developers.

## Application

This is where most dangerous security holes are found. Too often, developers are under high pressure to deliver programs on time. This means that their code is not thoroughly tested, and often contains security holes. Therefore, it is mandatory to have security mechanisms to encapsulate the application level. It must be impossible to bypass these mechanisms while trying to access the resources.

It is almost impossible to implement security without providing developers with tools and means through middleware software or the use of specialized libraries, with defined standardized APIs.

## Network Security

As the recent past shows, most kinds of attacks now occur from a distance, over a network. Networks themselves were not initially designed with security in mind. The vulnerabilities these attacks exploit can stem from several possible sources: the inherent design of the communication protocols, the design of the network topology, or the absence of a properly implemented quota (rate limiting) system to prevent denial of service attacks.

With the publicity surrounding the dangers of the *wide-open* Internet, many organizations concentrated all their network security efforts on the installation of firewalls around their networks. This approach results in heavily guarded network perimeters, while the inside of the networks have few security measures deployed. Analysis shows that most frauds are perpetrated by people from inside the organization, thus well inside the guarded network and not exposed to any check by the firewalls. Similarly, once an intruder breaches a firewall, through a hole in network design or a security flaw in the firewall itself, no barriers are left to prevent major damages.

Many security attacks easily pass through typical, level 3, firewalls. For example, Trojan horse software installed on the server maliciously or by error, can not be detected by a level 3 firewall. Content filtering firewalls may block such attacks but only when dealing with a known threat, which mandates frequent updates, and only if they can tackle the complete network throughput without being an unacceptable bottleneck.

## Discussion

An effective security implementation must be based on simultaneous secure mechanisms in all areas: hardware, software, and network infrastructure. In the end, the whole security process can be reduced to answers to the following questions:

- What level of security is needed?

- How to get it?

- How to know it is there?

- How to maintain it?

# Security Requirements for Carrier Class Servers

This section reviews carrier class systems, identifies traditional security practices, and then presents specific security-oriented requirements.

## Specificities of Carrier Class Systems

### Reliability

Reliability (as a part of availability) has long been emphasized for telecom systems (e.g., 5-nines, or 99.999%). In case of error or failure, it is seen as more desirable that the system continues to run and provide functionality in a fail-over fashion than not run at all. Unfortunately, fail-over modes of operations may have been designed in a way that is not necessarily fail-*safe* (i.e., safe in a security sense).

### Modular Multi-Server Systems

Architectures (e.g., wireless networks ones), use a variety of servers as basic building blocks (which may be implemented as clusters). These are interdependent and provide services to each other. In order for the whole system to work as intended, every server must function properly. They must trust each other. Security mechanisms can only be used within one server, or to make sure that it has secured connections to the right servers with which it has to exchange. One server cannot secure another.

### Remote Operation & Maintenance

As a consequence of a distributed architecture, O&M needs to be performed in a coordinated fashion. This implies remote O&M. The communication that supports this will need to be secured, whether they rely on a private dedicated network or on a public one.

**Context**

This section examines the typical characteristics of telecommunication servers from different points of view, while considering the advent of next generation telecom technology.

- Client machines: a wide range of devices from cellular phones and PDAs to desktop systems. The client's capacity to accept security mechanisms impacts server side security.

- Server machines: Powerful machines with huge amount of storage, memory and CPU, for instance a cluster with hundreds of nodes.

- Internal network or Private transport infrastructure, typically several internal IP based LANs used exclusively for the telecom servers, some of which for intra cluster communication.

- External network or Public transport infrastructure, typically shared among different users or service providers. It includes access mode mechanisms and external network. It provides IP connectivity for private or public networks (Gigabit Ethernet, ATM, SONET, Microwave), wired subscribers' lines (ADSL or coaxial cables), or wireless access.

- OS: most telephony equipment was running specialized operating systems developed internally. The current trend is to use more Components off the Shelf (COTS), running standard operating systems and protocols (e.g., POSIX and IP), especially for new services. This decreases the development costs and shortens the development time.

- Middleware: earlier systems relied on specialized internal solutions. Standard systems like CORBA are increasingly used.

- Application: There are two fundamental changes: introduction of new players from outside Telecom companies, and the introduction of new complex services. Various servers that are part of the telephony infrastructure (e.g., servers from SS7, 3G, WCDMA, and UMTS), have to support services that are new to it (e.g., mail and web), many of which allow complex interactions, including running code written by third parties (e.g., application servers supporting Java Beans). This changes how software is managed on new telecom servers, as telecom companies do not control the whole application development process: from inception to testing and maintenance.

For a long time, software development for Telecom systems was a field apart with its own peculiarities. It now follows many of the same concepts found in the IT business, like software reuse and using COTS. However, there are still many requirements specific to Telecom systems like fault tolerance and high availability.

The security mechanisms must be able to accommodate the typical context of hundreds of thousands of small nomadic clients (PDA or phone with limited memory and CPU power) to provide security for this high cumulative bandwidth in acceptable time delays. This may preclude complex computations on the clients, and requires servers capable of serving thousands of requests per second.

## Requirements

The security requirements for new Telecom applications are [Lang2000]:

- Support for many (potentially nomadic) users, multiple domains, multiple providers

- Support for dynamic hardware and software updates, especially urgent security updates

- Support for "multilateral security":

    o Secure interaction between domains with different security policies,

    o Dynamic security policies: ability to change the security policy (depending on server load, context, user privileges, new security threats, etc.) and make it effective rapidly.

- - Security Negotiation Facility

  - Many different parties involved, from different operators to different service providers

  - Many differing implementations and mechanisms used, including the best established standards (IPv4, IPv6, IPSec, ISAKMP, SSL/TLS, SSH, Radius, Diameter, etc.)

- Quality of Protection (QoP) depends on the type of service, and user. For example, one may not wish to implement costly security mechanisms for all clients (e.g., purchasing hardware keys for low volume clients).

- Secure Billing Facility:

- Non-repudiation (NR) service required

- Legal requirements: Data protection, legal data access,

- Security functionality for low performance/bandwidth systems

- Security for QoS services (e.g., real-time).

The security mechanisms implementing these requirements must in turn support:

- Fault tolerance, the security mechanisms must not become a single point of failure, on the other hand they must resist to possible failures of materials.

- High availability, the telecom applications are highly available. The security mechanisms corresponding to these applications must be reliable to the same degree in order to assume the security in all situations.

- Scalability:

- - Scalable cross-domain authentication

    - Scalable authorization

- Transparency:

    - Transparent security policy

    - Easy implementation of applications, it must not make the development difficult

    - Easy administration

    - Low performance overhead

- Consistency: coherent security policy must be applied among different components of servers and among involved servers in providing the service. This is to ensure an end-to-end security between clients and the server. For example, the same encryption algorithm must be supported on the server and clients

- Flexibility: each market may have specific requirements, and these evolve over time, even different clients of a same server may have different requirements

- Interoperability: between different security mechanisms for different operators, between different security technologies, protocols, and services

- Variety of quality of protection: providing the operator with different qualities of protection. This in order to allow the operator to implement different services according to the client needs.

## Security Level

Cluster-based telecom systems can be categorized according to the implicit trust or distrust of their components and environment: External network, Internal network, Server software (Operating System, Middleware, and Applications).

Security levels for clustered servers.

|         | Server application software | Internal network   | External network   |
|---------|-----------------------------|--------------------|--------------------|
| Level 0 | Implicitly trusted          | Implicitly trusted | Implicitly trusted |
| Level 1 | Implicitly trusted          | Implicitly trusted | Not trusted        |
| Level 2 | Implicitly trusted          | Not trusted        | Not trusted        |
| Level 3 | Not trusted                 | Implicitly trusted | Not trusted        |
| Level 4 | Not trusted                 | Not trusted        | Not trusted        |

In the table above, "not trusted" implies that mechanisms to prevent, detect, and respond to attacks must be implemented to fill the corresponding security void.

Level 0 corresponds to many existing user machines on the Internet. Level 1 and 2 are the case for many servers in enterprise environment, using a firewall for protection from attacks coming from the outside world. Level 3 security is a server capable of downloading code from a third party service provider through a dedicated line, (e.g., Java Enterprise Beans application server capable of importing third-party code); the internal network is trusted but not the application. Level 4 is the same server in a hostile environment where the internal network could be under attack.

Naturally Level 4 is the most secure level. However, the authors consider Level-2 security to be enough for a normal telecom server not exposed directly to clients and not in a hostile environment.

# Perspectives

Security incidents are more frequent than ever on the Internet, and attacks are of increasingly different types, from more varied sources, and are becoming more sophisticated. Meanwhile, there is a deep change in the Telecom market caused by the emergence of all-IP networks. Therefore, more efforts are being focused on security issues in the telecom world. One usual mistake is to base all the security on firewalls. As demonstrated many times recently, this is not enough to ensure security for a critical server. To increase security a step further, level 2 security can be attained today by combining different existing solutions. Even then, there are still two issues remaining.

First, the integration and management of these different solutions is very painful, when possible. This often results in the absence of interoperability between different security mechanisms. In many cases, the difficulties raised when integrating these systems result in an almost impossible system to maintain and upgrade. With time, many systems become obsolete or insecure by the difficulty of keeping up with numerous security patches and upgrades. Secondly, this does not seem to be enough in the coming years to protect key elements of a network.

Therefore, there is a need for coherent solutions implicating different aspects of security from packet filtering (firewalls) and intrusion detection, to access control and logging inside the operating system kernel. For the time being, no such solution is clearly emerging in the market.

Building a level-4 cluster requires implementing and putting together many security technologies that have been the focus of research for years, and turning this into a stable system that still has competitive performance characteristics. However, the authors are not aware of any real world deployment of such solutions, and this is most probably due to the lack of consensus on the solutions and their high management complexity.

Security protocols like IPSec are becoming a *de facto* standard for system security, just like SSL/TLS is for application security. Telecom servers must then provide support for them.

Also of crucial importance is to be able to quantify the impact of different security mechanisms on the performances and time delays for servers. This is a prerequisite to be able to propose adequate security measures to developers according to the application context. For example, what constitutes acceptable time delays differs for audio applications and image centric applications.

# Conclusions

As the number of attacks on the Internet increases everyday, security becomes a feature as essential as O&M, rather than luxury. In this perspective, the security vulnerability

February 2002

testing tends to become a complete part of IT activities as the conformance tests for network protocols.

Since it is not possible to eliminate 100% of the risks, much depends on the security level that is desired for servers. Finding the "right" trade-off between the desired level of security and the constraints due to performance issues is non-trivial. Security mechanisms with prohibitively high costs in performance are just as unrealistic as very speedy servers with no real security. The authors believe that Level 4 will be the desired level of security for many Telecom servers in the coming years, as a direct consequence of the requirements of forthcoming applications and environments.